# High-Speed Waveguide Integrated Silicon Photodetector on SiN-SOI Platform for Short Reach Datacom

AVIJIT CHATTERJEE[1], SAUMITRA[1,2], SUJIT KUMAR SIKDAR[1,2], AND SHANKAR KUMAR SELVARAJA[1*]

[1]*Center For Nano Science and Engineering, Indian Institute of Science, Bangalore*
[2]*Molecular Biophysics Unit, Indian Institute of Science, Bangalore*
*Corresponding author: shankarks@iisc.ac.in

**We present waveguide integrated high-speed Si photodetector integrated with silicon nitride (SiN) waveguide on SOI platform for short reach data communication in 850 nm wavelength band. We demonstrate a waveguide couple Si *pin* photodetector responsivity of 0.44 A/W at 25 V bias. The frequency response of the photodetector is evaluated by coupling of a femtosecond laser source through SiN grating coupler of the integrated photodetector. We estimate a 3dB bandwidth of 14 GHz at 20 V bias, highest reported bandwidth for a waveguide integrated Si photodetector. We also present detailed optoelectronic DC and AC characterisation of the fabricated devices. The demonstrated integrated photodetector could enable an integrated solution for scaling of short reach data communication and connectivity.**

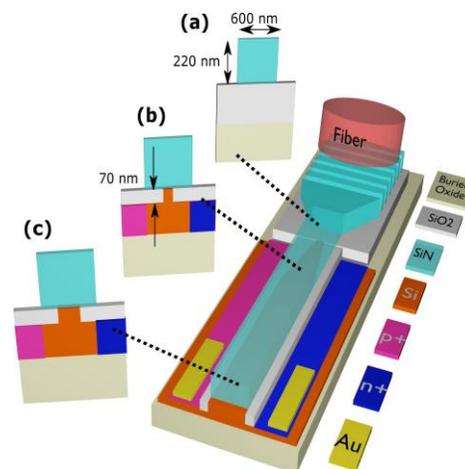

**Fig. 1.** Schematics of the proposed waveguide coupled Si photodetector. (a) SiN waveguide on $SiO_2$, (b) Tapered Si waveguide vertically coupled to SiN, and (c) Si *pin* photodetector with SiN waveguide.

Ever increasing data traffic drives the need for scalable, high speed, and energy efficient connectivity in the datacenters [1–3]. Optical connectivity offers an excellent solution to data connectivity bottleneck in datacenters. Widely used 850 nm VCSEL based Multimode fiber (MMF) technology is reaching maximum capacity, however, proposal to extend the MMF toward Wavelength Division Multiplexing (WDM) would scale up the capacity to 400 Gbps. In view of that, IEEE proposed 400 Gbps roadmap [4] for short reach rack-to-rack communication by 2020. Adapting WDM would require integration of wavelength multiplexers and demultiplexers, light sources and detectors in a single module. Integrating all the functional elements as a single photonic integrated circuit would enable a compact, cost-effective, and scalable platform.

It is essential that the circuit should be able to guide light with low-loss and is compact with the integration of light sources and detectors. Silicon (Si) photonics has demonstrated such a versatile platform in 1310 nm and 1550 nm wavelength band [5–7]. A similar strategy could enable a photonic circuit platform in 850 nm band as well. Recently, efforts are made to integrate III-V light sources and III-V/Graphene photodetectors on Silicon Nitride (SiN) waveguide platform [7, 8]. However, such hybrid integration requires complex assembly and post-processing technology, also the fabrication is not compatible with preferred Si process technology. It is desirable to reduce such non-standard integration. In this regards, Si is an excellent photodetector in the 850 nm band with higher responsivity than III-V photodetector. However, longer absorption length, RC time, and transit time in Si considerably limit bandwidth [9–11]. The bandwidth limitation of Si photodetector is significantly improved by realizing lateral *pin* photodetector in a guided wave configuration, particularly in a Silicon-on-Insulator (SOI) platform [12]. The thin Si in SOI has smaller depletion capacitance, which enhances the RC time limited bandwidth.

In this paper, we present SiN waveguide integrated high-speed Si *pin* photodetector on the SOI platform. Figure 1 illustrates a schematics of the proposed device that consists of SiN optical waveguide, SiN grating coupler, SiN-Si inverse taper coupler, and Si *pin* photodetector. For optical waveguide, SiN is an attractive CMOS compatible material that has large transparency window; visible to mid-IR wavelength band and explored widely for the linear and non-linear properties. [8, 13]

The light from an optical fiber is coupled into the chip using a grating fiber coupling in the SiN waveguide. The waveguide dimensions were optimized to support single mode operation with a waveguide width and thickness of 600 nm wide and 220 nm respectively. To couple 850 nm light from a fiber into the SiN waveguide, grating coupler and linear taper are used. An iterative optimization of the grating parameters such as grating period, etch depth and fill-factor were done to achieve maximum coupling efficiency. From the FDTD simulation, a coupling efficiency of 9.6 dB/coupler with a 1dB bandwidth of 12 nm for air-clad SiN grating couplers is achieved with a grating period of 650 nm and etch depth of 180 nm. The grating performance was experimentally verified by fabricating gratings in 220 nm thick LPCVD SiN over 2.2 $\mu m$ thick $SiO_2$. Figure 2 (a) and (b) shows the cross-section of the fabricated SiN grating coupler and SEM image of the SiN waveguide and grating coupler respectively. AFM image of the grating coupler is shown in Figure 2 (c) and (d) shows the etch depth of around 180 nm. The coupling efficiency is characterized using a fiber-coupled LED source (850-900 nm) and optical spectrum analyzer. Figure 2 (e) shows the measured coupling efficiency of a SiN grating coupler with 650 nm grating period yielding a maximum coupling efficiency of 12.6 dB/coupler around 867 nm.

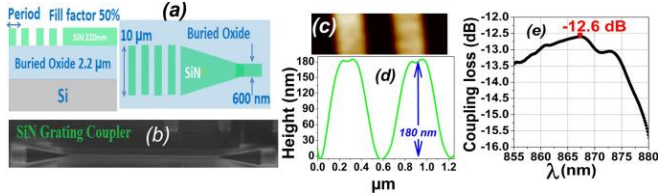

**Fig. 2.** (a) Schematics of the test structure (top and cross section view), (b) SEM image of the fabricated SiN grating, (c) AFM image of the grating, (d) Shows the etch depth of 180 nm of gratings obtained from AFM, (e) Grating coupling efficiency Vs wavelength

The propagating light in the SiN layer should be coupled into the underlying Si waveguide photodetector layer. In the proposed structure, the evanescent mode coupling is used to couple light between the SiN and Si layer. Due to a large mode mismatch difference in the two waveguides, the vertical coupler requires a longer length to efficiently couple light from SiN to Si waveguide. However, longer coupling length would increase the Si photodetector length thus degrades the 3dB bandwidth due to higher depletion capacitance [14]. To achieve higher coupling efficiency with a large bandwidth of the photodetector, we introduce Si inverse taper as illustrated in Figure 1. Si taper helps to achieve higher coupling efficiency between SiN and Si with shorter overlapping coupler length due to better phase matching. Shorter overlapping length enhances the bandwidth of the photodetector as well. To optimize the geometry of the Si inverse taper, we use commercial simulation software Fimmprop [15] and Lumerical FDTD [16]. Figure 3a shows cross-section beam propagation and mode coupling from the top SiN layer into Si layer. Successive optimization of the vertical direction coupler length and taper parameters were performed to achieve maximum coupling. Figure 3b shows the simulated coupling efficiency versus taper length ($L_T$) at 850 nm wavelength for optimized silicon taper as described in Figure 3c. An optimised coupling efficiency of 72% is obtained from the simulation for a 15 $\mu m$ long Si taper. The coupling efficiency could be increased by making the longer direction coupler, however, this could result in lower bandwidth. Therefore, Si taper length of 15 $\mu m$ is optimum to achieve high responsivity as well as large bandwidth of the photodetector. Si absorbs the coupled light in the 850 nm wavelength band and generates Electron and Hole Pairs (EHPs). The generated EHPs are collected laterally by the photodetector along the propagation direction. Since *pin* photodetector has larger absorption region and lower depletion capacitance, we have chosen *pin* over *pn* junction photodetector. Moreover, the shallowly etched ridge geometry of the Si photodetector confines the coupled light in the intrinsic region which minimises the photocarrier generation outside the depletion region. Photocarriers outside depletion region must diffuse before getting collected, which limits the bandwidth of the photodetector. Hence, ridge waveguide helps to improve the photodetector bandwidth further. Furthermore, to achieve the highest bandwidth, intrinsic width of the *pin* needs to be optimised. From device simulations, intrinsic width of 200 nm for 15 $\mu m$ long *pin* photodetector gave a maximum 3dB bandwidth of 25 GHz.

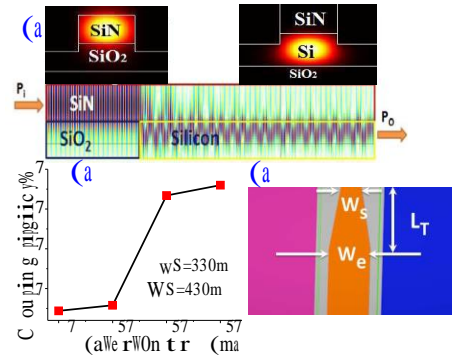

**Fig. 3.** (a) Optical simulation of Si and SiN waveguide and coupler, (b) Coupling efficiency from SiN waveguide into Si ($P_O/P_i$) for different Silicon taper length ($L_T$), and (c) Top View of the Si taper configuration.

The optimized devices were fabricated on a photonics SOI substrate having 220 nm thick Si device layer on 2 $\mu m$ thick buried oxides. Figure 4 depicts an overview of the fabrication process flow. The photodetector structure in the Si layer was patterned by using optical lithography followed by dry etching. The Si pattern was then coved with silicon dioxide and planarized using CMP. The doping of $p+$ and $n+$ regions were realized using diffusion doping process. The doping windows were placed 4 $\mu m$ apart to avoid overlapping of the dopants during diffusion doping process [17]. To form the SiN waveguide and grating coupler, LPCVD SiN was deposited and patterned using e-beam lithography, and dry etch process. Before $SIN$ deposition 50 nm $SiO_2$ is deposited as a liner layer to project the underlying Si during $SiN$ pattering. Finally, Chrome/Gold electrical contacts were made on $p+$ and $n+$ dopant window. Figure 5 shows SEM image of the fabricated device. A set of three devices with varying Si taper parameter were fabricated. Figure 5 shows the set of devices (D1, D2 and D3) that were fabricated. All the devices were designed with a detector length of 100 $\mu m$. Since the junction starts from the taper, the effective detector length is $L_T + 100 \mu m$.

Experimental validation of the fabricated device was performed by measuring responsivity and bandwidth of the pho-todetector. Initial DC characterization is performed to determine the dark current and responsivity. Light is coupled into the chip using a grating coupler as mentioned earlier. For responsivity measurement, the optical power delivered into the waveguide is measured using a reference SiN waveguide with in and out coupler. Figure 6 (a) and (b) show the measured dark current and photocurrent of the fabricated devices (D1, D2 and D3) each with 115 $\mu m$ length. All the device configurations measured a photocurrent that is two orders of magnitude higher than the dark current. We measure a dark current as low as 46 nA at 10 V bias, at the same voltage we measure a photocurrent of 5.09 $\mu A$ for device D3. We measure the highest photocurrent of 7.5 $\mu A$ at 25 V for device D3. At higher bias voltage photocurrent increases due to impact ionization. From Figure 6(b) we observe that the photocurrent saturates beyond 15 V for device D3. This occurs due to space charge effect which reduces the electric field at the center of the *pin* [12]. This effect is more prominent for device D3 than D1 and D2 since D3 has higher dark current than D1 and D2 at higher bias voltages.

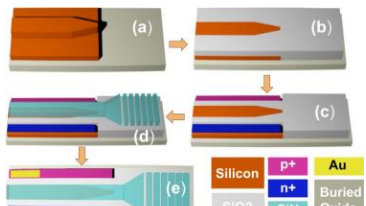

**Fig. 4.** Fabrication process flow (a) Silicon patterning using optical lithography and dry etching, (b) $SiO_2$ deposition and planarization, (c) p+ and n+ junction region defined using diffusion doping, (d) 220 nm LPCVD SiN deposited and patterned (waveguide and grating coupler) using e-beam lithography and dry etching, and (e) Cr-Au deposited by e-beam evaporator followed by lift-off to define electrical contacts.

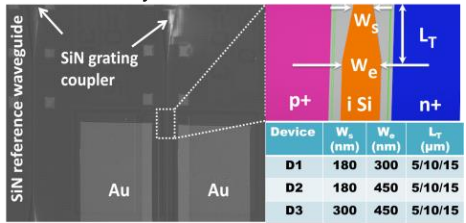

**Fig. 5.** Scanning electron microscope image of the fabricated device (left) and Taper parameters of Device D1, D2, and D3 fabricated (device total length = $L_T + 100 \mu m$), intrinsic width = $4 \mu m$ (right).

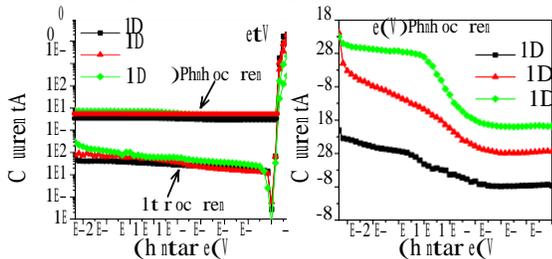

**Fig. 6.** IV characteristics of photodetector for different devices (D1-D3) of length L = 115 $\mu m$, (a) Log scale, (b) Linear scale

The responsivity of the detector is measured from the DC photocurrent and power coupled into the waveguide. Figure 7 shows the summary of the responsivity of the fabricated detectors. Irrespective of the Si-SiN taper design, detector with longest taper length of 15 $\mu m$ measured maximum responsivity in each design set. As the length of the total device increases, the responsivity increases due to better power coupling between SiN and Si waveguide. However, responsivity for different devices (D1-D3) are almost identical up to 10 V bias due to insignificant change in absorption region. Beyond 10 V bias, responsivity of (D1-D3) starts to deviate and shows increase in responsivity with bias due to impact ionization effect. we measure a maximum responsivity of 0.44 A/W at 25 V for device D3.

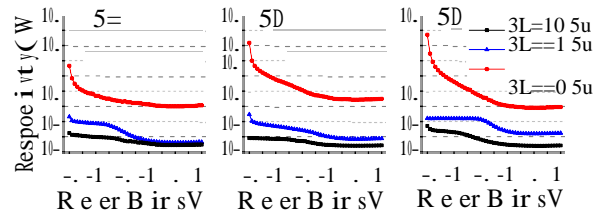

**Fig. 7.** Responsivity of devices D1, D2, D3 for different length of Si photodetector at various bias voltage.

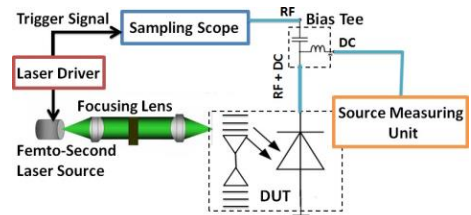

**Fig. 8.** Frequency response measurement setup with femtosecond pulse Laser.

The frequency response of the detector is measured using a femtosecond pulsed laser source [18]. The 800 nm femtosecond laser source generates a pulse of width 280 fs, 200 mW of average power at 80 MHz repetition rate. Figure 8 shows a schematic of the experimental setup used to measure the pulse response of the detector. Generated current pulse output is fed into the Bias Tee (26 GHz cut off) to separate the AC and the DC components. DC component is measured by a source measuring unit. The AC component of the pulse output is measured using a Wide-Bandwidth Oscilloscope (DCA-X 86100D).

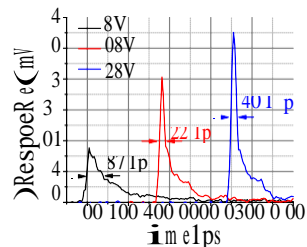 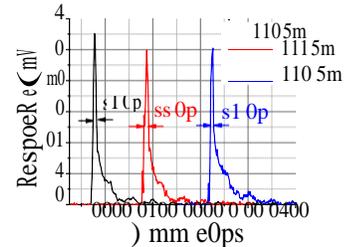

**Fig. 9.** Pulse response of the photodetector (device D3) at different reverse biases.

**Fig. 10.** Pulse response of photodetector for different lengths of device D3 at 20 V bias.

Similar to DC measurement, we measured both waveguide coupled and direct illumination frequency response. The femtosecond source of spot size 5 $\mu m$ is illuminated on top of the intrinsic region to evaluate the frequency response of the detector. Figure 9 shows the temporal pulse response of device D3 with 105 $\mu m$ length for different bias voltages. The 3dB frequency ($f_{3dB}$) of the photodetector is estimated from the full width at half maximum (FWHM) of the current pulse by the relation $f_{3dB} = \frac{0.45}{FWHM}$[19]. We measured the best FWHM of 29 ps corresponds to the 3dB bandwidth of 15.5 GHz at 10 V reverse bias. FWHM at 0 V is wider than at 10 V due to the longer transit time of the carriers at lower bias voltages. However, with an increase in the bias voltage to reduce transit time could result in avalanche subsequently reducing the detector speed. Figure 9 depicts the evolution of the photodetector response with varying bias. Figure 10 shows the pulse output for different lengths of the photodetector. FWHM does not vary significantly with the change in device length that suggests detector bandwidth is not RC time-limited rather transit time limited. To confirm transit time limitation, depletion capacitance, contact pad resistance, and diode series resistance of the *pin* were measured. Depletion capacitance of the *pin* is estimated from the relation $C = \frac{e_0 e_{Si} A}{W_{Si}}$, where $A$ is the cross-section area of the photodetector ($100 \mu m \times 220 nm$) and $W_{Si}$ ($3 \mu m$) is the intrinsic width of the *pin* which is derived by subtracting the lateral diffusion length of dopant from the intrinsic width designed in lithography ($4 \mu m$). Calculated capacitance for the *pin* is extremely low (0.76 fF). Calculated Contact pad capacitance is in the order of sub fF hence neglected. Contact pad resistance and diode series resistance are obtained from the I-V of the photodiode that comes around 416 Ω. It is evident from the low-capacitance and resistance the device bandwidth is transit time limited.

As mentioned earlier, femo-second source is waveguide coupled into the detector through the grating coupler. Since the source was placed vertical to the grating coupler instead of the designed angle of $10^o$, the power coupling was lower, however, sufficient to measure the pulse response. Figure 11 and 12 present the photodetector response obtained by coupling the femtosecond laser through the SiN grating coupler of the integrated photodetector. Figure 11 shows the pulse response for different devices (D1-D3). Figure 12 summarizes the frequency response of device D3 with different device lengths. Changes in device length of the detector does not cause a significant change in frequency response because the detector bandwidth is transit time limited as discussed in the previous section. Best FWHM of 32 ps corresponds to 14 GHz 3dB bandwidth at 20 V reverse bias is obtained for device D3 with 115 $\mu m$ length. Since femtosecond laser source is illuminated on the SiN gratings at zero degree angle, the coupling efficiency of SiN grating is low. Due to which amplitude of the pulse output at lower reverse bias voltages falls under the noise floor of the measuring instruments.

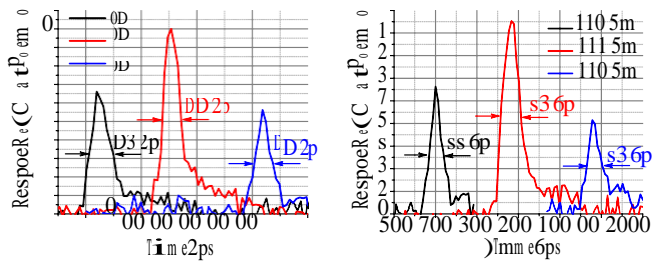

**Fig. 11.** Pulse response for devices D1-D3 of length 105 $\mu m$ at 20 V bias with pulse laser coupled to SiN grating coupler.

**Fig. 12.** Effect of device length (device D3) on pulse response at 20 V bias with pulse laser coupled to SiN grating coupler.

| Parameters | [11] | [12] | [14] | [7] | This Work | This Work |
|---|---|---|---|---|---|---|
| Dark Current (nA) | 0.06 | 2X10³ | 0.23 | 22 | 46 | 75 |
| Responsivity (A/W) | - | 0.3 | 0.32 | 0.12 | 0.2 | 0.29 |
| Bandwidth (GHz) | 19.5 | 16.4 | - | 20 | 15.5 | 14 |
| Bias (V) | 5 | 14 | 9 | 2 | 10 | 20 |
| Integrated | No | No | Yes | Yes | No | Yes |
| Platform | Si | SOI | Si | III-V | SOI | SOI |

**Table 1.** Comparison with the integrated photodetectors for short reach optical interconnects in the literature.

Table 1 compares the results of this work with previously published works in literature. The bandwidth of 14 GHz and responsivity of 0.29 A/W with 75 nA dark current at 20 V reverse bias are obtained with 115 $\mu m$ long device. To the best of our knowledge, this is the highest reported 3dB bandwidth for waveguide integrated Si photodetector.

In conclusion, we have successfully demonstrated an integrated lateral silicon *pin* photodetector in SiN-on-SOI platform. Silicon *pin* is realized using simple diffusion doping. SiN waveguides modes are coupled into the intrinsic region by an optimized inverse taper vertical coupler. We obtained the highest responsivity of 0.44 A/W at 25 V, and to the best of our knowledge highest 3dB cut off frequency of 14 GHz at 20 V bias for a waveguide integrated silicon photodetector. We find that bandwidth increase with increasing bias voltage reveals that the speed is limited by transit time which can be further improved by the narrower intrinsic width of the *pin* junction that can be achieved using implantation doping. The demonstration shows a feasible method to integrated standard Si process and SiN integration to achieved high-speed photodetector operating in short reach wavelength band. The demonstration allows integration of wavelength selective detection that can be used for short reach wavelength division multiplexing.

**Acknowledgement**. "We thank DST-SERB for funding this research. We also acknowledge funding support from MHRD through NIEIN project, from MeitY and DST through NNetRA"